\documentclass[aps,prx,twocolumn]{revtex4-2}
\usepackage{graphicx,amsmath,amsfonts,amssymb}
\usepackage{bm}
\usepackage[mathlines]{lineno}
\usepackage[normalem]{ulem} 

\usepackage[dvipsnames]{xcolor}
\usepackage{ulem}

\usepackage{soul}

\newcommand{\tr}{\operatorname{tr}}

\usepackage{xcolor,hyperref}
\hypersetup{
   colorlinks,
   linkcolor={blue!50!black},
   citecolor={blue!50!black},
   urlcolor={blue!80!black}
}

\renewcommand{\=}{\!=\!}

\begin{document}

\title{Size selection of crack front defects:\\ Multiple fracture-plane interactions and intrinsic lengthscales}
\author{Meng Wang$^1$}
\author{Eran Bouchbinder$^2$}
\author{Jay Fineberg$^1$}
\affiliation{$^1$The Racah Institute of Physics, The Hebrew University of Jerusalem, Jerusalem, 91904, Israel\\
$^2$Chemical and Biological Physics Department, Weizmann Institute of Science, Rehovot 7610001, Israel}

\begin{abstract}
Material failure is mediated by the propagation of cracks, which in realistic 3D materials typically involve multiple coexisting fracture planes. Multiple fracture-plane interactions create poorly understood out-of-plane crack structures, such as step defects on tensile fracture surfaces. Steps form once a slowly moving, distorted crack front segments into disconnected overlapping fracture planes separated by a stabilizing distance $h_{\rm max}$. Our experiments on numerous brittle hydrogels reveal that $h_{\rm max}$ varies linearly with both a nonlinear elastic length $\Gamma(v)/\mu$ and a dissipation length $\xi$. Here, $\Gamma(v)$ is the measured crack velocity $v$-dependent fracture energy and $\mu$ is the shear modulus. These intrinsic lengthscales point the way to a fundamental understanding of multiple-crack interactions in 3D that lead to the formation of stable out-of-plane fracture structures.
\end{abstract}
\maketitle

\begin{figure}[ht!]
	\includegraphics[width=1\linewidth]{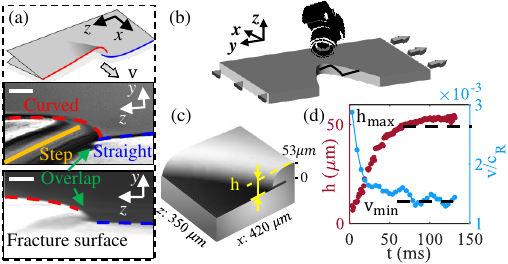}
\caption{Fracture experiments on steps formed by segmented crack front interactions in polyacrylamide gels. (a) (top) Step structure; steps are composed of straight (blue) and curved (red) segments that overlap in the $x\!-\!z$ plane (a right-handed Cartesian coordinate system $x\!-\!y\!-\!z$ is consistently used throughout the figure). (center) Microscopic photograph of an arrested crack front with a step, imaged through the transparent gel. Light is reflected from the lower fracture surface into the microscope. Both fracture planes are clearly visible (colors as in the top panel). The yellow curve traces the step-line.  (Bottom) A second step, enhanced with overexposed lighting. Green arrows denote overlapping regions. Cracks propagate in $x$. Scale bars: 50 $\mu$m. (b) Schematic of experimental set-up used to image (a). Step-lines are marked by black lines. (c) Measured step on a fracture surface by profilometer. The out-of-plane height, $h$, is the separation in $y$ of the intersecting fracture planes. (d) Typical evolution of $h(t)$ and $v(t)$ during step formation. Steps grow before stabilizing at an asymptotic height, $h\!\to\!h_{\rm max}(v)$ as $v\!\to\!v_{\rm min}$.}
\label{fig:fig1}
\end{figure}

Materials fail by the propagation of cracks. The structure and dynamics of `simple cracks' in brittle materials are well-described by the classical theory of cracks, Linear Elastic Fracture Mechanics (LEFM)~\cite{freund1998dynamic,broberg1999cracks}. We define a `simple crack' as a smooth straight line left behind a propagating tip in an effectively 2D medium. A simple crack produces a scale-free singular stress field $\sim\!1/\sqrt{r}$, as a function of the distance $r$ from its tip. In general, real materials are not 2D, hence there are no true 1D cracks. Yet, if the leading edge of a crack is a smooth, straight line in the thickness direction $z$, normal to its propagation direction $x$, simple cracks correspond to smooth $x\!-\!z$ planes, whose edge forms a `singular front' (see Fig.~\ref{fig:fig1} for a definition of the coordinate system). When a crack front is translationally invariant along $z$, the 2D LEFM description is valid and fracture becomes an effectively 2D problem.

Even under these conditions there are two tacit assumptions: the material surrounding the crack tip is linearly elastic down to the smallest $r$ and all dissipative processes in fracture are confined to a point-like region at $r\rightarrow 0$. Breaking these assumptions defines two lengthscales, a nonlinear elastic length $\ell_{\rm nl}\!\sim\!\Gamma(v)/\mu$ and a dissipation length $\xi$~\cite{livne.08,bouchbinder2008weakly,bouchbinder20091,livne2010,bouchbinder.09b}. $\ell_{\rm nl}$ is the scale where nonlinear elasticity significantly affects the deformation around the crack front. $\ell_{\rm nl}$ is proportional to the ratio of the fracture energy $\Gamma(v)$, which generally depends on the crack propagation velocity $v$, and the shear modulus, $\mu$~\cite{livne.08,bouchbinder2008weakly,bouchbinder20091,livne2010,bouchbinder.09b}.  Dissipation occurs at the scale $\xi$. Both of these `internal' scales break the scale invariance of the $1/\sqrt{r}$ field \cite{livne.08,livne2010,bouchbinder2010dynamics,bouchbinder2014dynamics,long2021fracture,bouchbinder.09b}. In tensile (mode-I) fracture, where loading is applied normal to the crack plane (in the $y$ direction), these scales play critical roles in determining the near-tip structure and stability of 1D cracks~\cite{livne.08,livne2010,bouchbinder20091,bouchbinder2008weakly,bouchbinder2010dynamics,goldman2012intrinsic, bouchbinder2014dynamics,long2021fracture,qi2019mapping,chen2017instability,lubomirsky2018universality,vasudevan2021oscillatory,bouchbinder.09b}.

Once translational invariance in $z$ is lost, cracks become decidedly more complex, featuring both in- and out-of-plane structures~\cite{fineberg.99,lubomirsky2023quenched,sommer1969formation,lazarus2008comparison,leblond2011theoretical, pons2010helical,chen2015crack,pham2016growth,leblond2019configurational,vasudevan2020configurational,pham2017formation,pons2010helical,chen2015crack,ronsin2014crack,lubomirsky2024facet,tanaka1998discontinuous, baumberger2008magic,kolvin2018topological,wang2022hidden,wang2023dynamics,steinhardt2022material,ronsin2014crack,lubomirsky2024facet,xue2024elastomers,wei2024complexity}. This may result from dynamic symmetry-breaking instabilities at medium to high $v$ ~\cite{fineberg.99,lubomirsky2023quenched}. In the slow fracture regime --- our focus here --- invariance may be lost due to mode-mixity, i.e., remote tensile loading (mode-I) accompanied by a small shear along $z$ (anti-plane shear, mode-III)~\cite{sommer1969formation}. Extensive theoretical and experimental work on mixed-mode I+III fracture~\cite{lazarus2008comparison,leblond2011theoretical, pons2010helical,chen2015crack,pham2016growth,leblond2019configurational,vasudevan2020configurational,pham2017formation,pons2010helical,chen2015crack,ronsin2014crack,lubomirsky2024facet}
showed that planar crack fronts transition to arrays of tilted segments under these conditions. These segments coarsen with time until only few such fragmented sections remain, each forming its own fracture plane. Propagating step-like structures, `steps', on  fracture surfaces mark the intersection of these planes. Steps also occur in nominally mode-I fracture~\cite{tanaka1998discontinuous, baumberger2008magic,kolvin2018topological,wang2022hidden,wang2023dynamics,steinhardt2022material,ronsin2014crack,lubomirsky2024facet}, emerging from the coupled effect of finite quenched disorder and mesoscopic mode-III fluctuations~\cite{lubomirsky2024facet}. Close examination of steps in elastomers~\cite{tanaka1998discontinuous, baumberger2008magic,kolvin2018topological,wang2022hidden,wang2023dynamics} has revealed that they are actually complex 3D topological structures, as shown in Fig.~\ref{fig:fig1}a.

At a step, two crack fronts within different $x\!-\!z$ planes separated in $y$ by a distance $h$, intersect. These fracture planes meet at two disconnected and slightly overlapping front segments, as highlighted by the blue and red lines in Fig.~\ref{fig:fig1}a. The leading front in $x$ creates a flat fracture surface in its wake. The second front lags behind the first and curves in the $y\!-\!z$ plane towards the flat plane already created by the leading front. This curved region always overlaps (in $z$) the flat plane. Experimentally, steps are topologically constrained, where loss of stability occurs only when their topology is momentarily broken~\cite{kolvin2018topological,wang2022hidden,wang2023dynamics}.

Once formed, steps can either grow or decay ~\cite{kolvin2018topological, steinhardt2023geometric}. Their separation distance $h$ evolves (Fig. \ref{fig:fig1}c,d) until stabilizing at $h_{\rm max}$, the `step size'. Little is known about the 3D interaction between the two fronts, and in particular about how their interaction selects $h_{\rm max}$. Characterizing $h_{\rm max}$ and its scaling properties is an important first step in understanding the nontrivial interactions between different fracture planes that both create these unique, yet ubiquitous, out-of-plane structures, and prevent them from merging/collapsing into a single fracture plane. Here, we quantify $h_{\rm max}$ in a variety of different materials and show how $h_{\rm max}$ is determined by nonlinear interactions between the two fronts, involving both intrinsic lengthscales $\ell_{\rm nl}$ and $\xi$.

Our experiments utilized numerous polyacrylamide hydrogel samples composed of acrylamide monomers crosslinked with bis-acrylamide. These incompressible brittle gels have been widely used to study material failure, both in the context of highly-deformable soft materials (e.g.,~\cite{long2021fracture}) and as representatives of a broader class of brittle materials~\cite{bouchbinder2010dynamics, bouchbinder2014dynamics}. As illustrated in Fig.~\ref{fig:fig1}b, gel samples with dimensions $x\times y \times z\=50 \times 30 \times 1$ mm$^3$ were loaded under uniaxial tension in $y$. We controlled the nominally mode-I crack dynamics through the stored strain energy prior to crack initiation. Each sample was first loaded to a desired applied stretch, $1.04<\!\lambda\!<\!1.18$. We then initiated crack propagation in $x$ by inserting a small cut on the mid-plane at a sample edge. Crack propagation was captured using a fast $1920\!\times\!1080$ pixel camera, mounted above the sample, with 4.25--15.7 $\mu$m/pixel spatial and 200--5000 Hz temporal resolutions. The crack fronts' mean velocity, $v$, was measured using the centroids of the caustics surrounding each crack front. (Caustics are caused by extreme deformation in $z$ in the near-front region~\cite{wang2022hidden}). We focus on slow crack propagation $v\!<\!0.15\,c_{_{\rm R}}$, where $c_{_{\rm R}}$ is the Rayleigh wave-speed of each material.

At low velocities, front segmentation forms steps that nucleate from either the rough initial cut or interactions with local material heterogeneities~\cite{lubomirsky2024facet}. While heterogeneities can nucleate steps, they have no effect on $h$ beyond their characteristic size~\cite{steinhardt2022material,lubomirsky2024facet, Supplementary}. Steps move along the propagating crack front in the direction that lengthens (shortens) the curved (straight) segment. Steps form faceted fracture surfaces by imprinting, through their motion, step-like lines, `step-lines'~\cite{tanaka1998discontinuous,kolvin2018topological}. We determined the step height, $h[v(t)]$, through post-mortem measurements of the step-lines on fracture surfaces using a white-light profilometer (Fig.~\ref{fig:fig1}c). For each measurement position along a step-line, occurring at time $t$, we also measured $v(t)$ (see below). $h(t)$ quantifies the separation distance (in $y$) of the interacting fracture planes forming the steps within the material (undeformed) frame~\cite{kolvin2018topological}.
\begin{figure}[ht!]
	\includegraphics[width=1\linewidth]{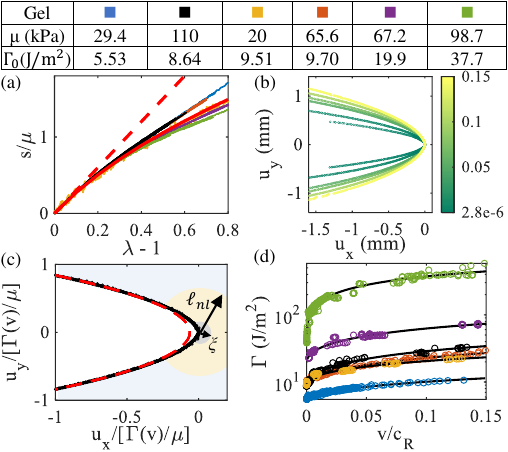}
	\caption{(Top) Table of $\mu$ and $\Gamma_0$ for the 6 gels used. (a) Nominal tensile stress normalized by the shear modulus, $s/\mu$, vs.~strain, $\lambda-1$, for all 6 gels. Dashed and solid red lines: linear elastic and neo-Hookean constitutive relations, respectively. The normalized stress-strain curves of blue, red and black gels perfectly overlap. (b) CTOD measurements of the orange gel at different $v$, color bar: $v/c_{_{\rm R}}$. (c) All CTOD in (b) collapse (black), when scaled by $\Gamma(v)/\mu$.  CTODs deviate from the LEFM prediction (red dashed parabola) within the nonlinear elastic region near the crack tip, $\ell_{\rm nl}$. The dissipation length $\xi$ is schematically shown. (d) Fracture energies $\Gamma$ vs.~$v/c_{_{\rm R}}$ for all gels. $\Gamma (v)$ are well described by $\Gamma_0[1 + a(v/c_{_{\rm R}})^b]$ (black curves)~\cite{Supplementary}.}
\label{fig:fig2}
\end{figure}

We focus on crack fronts forming a single step-line. In this configuration, the crack velocity at each step is approximately the mean velocity $v$ determined using caustics~\cite{wang2022hidden}. Figure~\ref{fig:fig1}d shows a typical example of both $h(t)$ and the resultant $v(t)$. Upon nucleation, $h$ increases from a few $\mu$m to a $v$-dependent asymptotic value, $h_{\rm max}$. In satisfying energy balance, step growth (producing extra fracture surface) causes $v$ to decrease towards its asymptotic value, $v\!\to\!v_{\rm min}$, as $h\!\to\!h_{\rm max}(v_{\rm min})$~\cite{wang2022hidden,wang2023dynamics}.

How is $h_{\rm max}$ selected? Emergent lengths in 2D crack dynamics, e.g., high-speed oscillatory instability wavelengths~\cite{livne.07} and a length related to a quasistatic macro-branching instability~\cite{baumberger2010convective}, are governed by either the nonlinear length, $\ell_{\rm nl}$, the dissipation length, $\xi$, or both~\cite{bouchbinder20091,bouchbinder2008weakly,qi2019mapping,chen2017instability,lubomirsky2018universality,vasudevan2021oscillatory}. $\ell_{\rm nl}$ has also been shown to be related to the cross-hatching and \'echelon instabilities in soft materials~\cite{baumberger2008magic,ronsin2014crack}. Are the 3D crack interactions producing $h_{\rm max}$ also governed by these intrinsic scales? To address this question, we used 6 different gels having a wide variety of mechanical properties controlled by the monomer-to-crosslinker molar ratio, $M$, and the total polymer concentration $\rho_{\rm g}$ in the range $13.8\%\!<\!\rho_{\rm g}\!<\!40.6\%$. The blue, orange and black gels in Fig.~\ref{fig:fig2} have a fixed $M$, with different $\rho_{\rm g}$. The remainder differ in both $M$ and $\rho_{\rm g}$~\cite{Supplementary}.

We first characterize the gels' mechanical response in uniaxial tension. Figure~\ref{fig:fig2}a shows that all gels follow linear elasticity (red dashed line) for small strains, $\lambda-1\!\to\!0$. For moderate (up to $\sim\!30\%$) strains, all follow the nonlinear elasticity described by the incompressible neo-Hookean law~\cite{holzapfel,Supplementary} (red solid line), despite their wide variation of shear moduli, $\mu$. At large strains ($\sim\!60\%$), the gels slightly deviate from neo-Hookean elasticity.

We also characterized the $v$ dependence of all fracture energies $\Gamma(v)$, using the crack tip opening displacement (CTOD) of simple cracks, as in~\cite{boue2015failing}. The method accounts for applied background stretches, and agrees well with J-integral measurements for slow fracture under moderate stretch levels~\cite{boue2015failing}. Fig.~\ref{fig:fig2}b presents the CTOD of one gel at different $v$. All CTOD collapse onto a single shape (Fig.~\ref{fig:fig2}c), once normalized by the nonlinear scale $\Gamma(v)/\mu$~\cite{goldman2012intrinsic}. The collapse at small scales, where the CTOD deviates from the LEFM prediction (the red parabola), indicates the existence of a nonlinear elastic zone of size $\ell_{\rm nl}\!\sim\!\Gamma(v)/\mu$, surrounding a dissipation zone of size $\xi$ around the crack front. So long as $\xi$ is small, measurements of $\Gamma$ using the CTOD are valid~\cite{boue2015failing}.

In Fig.~\ref{fig:fig2}d, we present the measured $\Gamma(v)$ for $0\!<\!v/c_{_{\rm R}}\!<\!0.15$. A wide variation of $\Gamma(v)$, ranging from few to hundreds of J/m$^2$, is observed. For each gel, $\Gamma(v)$ is well-described by the nonlinear function, $\Gamma(v)\= \Gamma_0[1 + a(v/c_{_{\rm R}})^b]$. $\Gamma_0\!\equiv\!\Gamma(v\!\to\!0)$ is determined through best fits to the experimental data ($a$ and $b$ are given in~\cite{Supplementary}). Such power-law dependence of $\Gamma(v)$ exists in a wide range of polymers, including PMMA~\cite{maugis1985subcritical}. Moreover, $\Gamma (v)$ is nearly the same (similar values of $b$) for gels with fixed $M$. $b$ decreases for more highly entangled gels (larger $M$), in agreement with recent experiments showing that highly entangled polymer networks produce gels with extreme fracture energies~\cite{kim2021fracture}.

\begin{figure}[h]
	\includegraphics[width=1\linewidth]{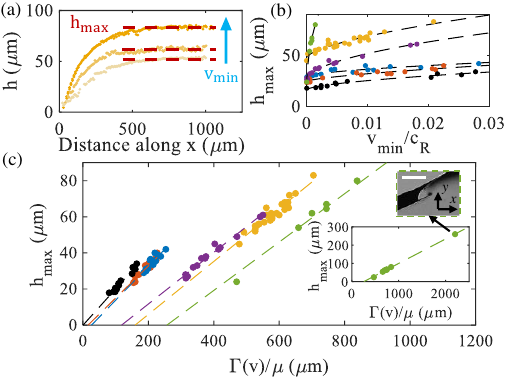}
	\caption{(a) $h$ vs.~step propagation distance in $x$ for the orange gel in Fig.~\ref{fig:fig2}, for asymptotic velocities $v_{\rm min}/c_{_{\rm R}}\!=\!0.0014, 0.0063, 0.021$ (bottom to top, see arrow). (b) $h_{\rm max}$ vs.~$v_{\rm min}/c_{_{\rm R}}$ (colors as in Fig.~\ref{fig:fig2}). Black dashed curves denote the best power-law fits, see text and~\cite{Supplementary}. (c) $h_{\rm max}$ vs.~$\Gamma(v)/\mu$. Dashed lines are the best linear fits. Inset: Data for green gel in the main panel extends over one decade. The data point for a step with the largest $h_{\rm max}$ is shown in the  photograph of the crack surface. Scale bar: $500\,\mu$m.}
\label{fig:fig3}
\end{figure}
Numerous experiments in all of the gels provided $h_{\rm max}(v)$. Three typical examples of $h(x)$ for a given gel, at different $v$ (different applied $\lambda$) are given in Fig.~\ref{fig:fig3}a. Upon step formation, $h$ increases until asymptotically reaching  $h_{\rm max}(v)$, which increases with $v_{\rm min}$. $h_{\rm max}(v_{\rm min})$ is presented for all 6 gels in Fig.~\ref{fig:fig3}b. For each gel, $h_{\rm max}$ varies nonlinearly with $v_{\rm min}$ \cite{Supplementary}.

Is the nonlinear behavior of $h_{\rm max}$ inherited from that of $\Gamma(v)$ through $\ell_{\rm nl}(v)$? To test this, we plot $h_{\rm max}$ vs.~$\Gamma(v)/\mu$ in Fig.~\ref{fig:fig3}c. Remarkably,  $h_{\rm max}$ is a well-defined {\it linear} function of $\Gamma(v)/\mu$ (dashed lines) for all gels, as originally suggested and anticipated in~\cite{baumberger2008magic,ronsin2014crack}. This strongly suggests the importance of the nonlinear scale, $\ell_{\rm nl}\!\sim\!\Gamma(v)/\mu$ in multi-crack interactions. The aforementioned linear function corresponds to
\begin{equation}
	h_{\rm max} = \alpha\,(\Gamma(v)/\mu - \ell_0) \ ,
	\label{eq:h_lnl}
\end{equation}
where the slope $\alpha$ is nearly identical for {\it all} of the gels. The approximate material {\it independence} of $\alpha$ in Eq.~\eqref{eq:h_lnl}, therefore, suggests that the interaction between crack front segments is governed by the nonlinear elastic scale $\ell_{\rm nl}$, which, in turn, may govern the stability of segmented crack fronts. This is a major finding.

What determines $\alpha$? The nearly singular material deformation near the crack front implies that front interactions take place at high local stretch levels~\cite{wei2024complexity}. Up to $\sim\!80\%$ strain, our uniaxial tension measurements in Fig.~\ref{fig:fig2}a indicate that all gels predominantly follow neo-Hookean elasticity, with small, yet systematically increasing deviations at large values of $\lambda$. The {\it nearly} neo-Hookean response may account for the predominantly material-independent $\alpha$, while the deviations may be related to non-neo-Hookean elastic nonlinearity.

To quantify the material deviations from incompressible neo-Hookean elasticity, we follow~\cite{knowles1983large} and express the 2D plane-stress energy functional of incompressible gels as
\begin{equation}
	{U(I, J)} = \frac{1}{2}\mu(\mathring{I_1}-3) + \epsilon\,\mu(\mathring{I_1} - \mathring{I_2}) \ .
\label{eq:Energy_den}
\end{equation}
Here, $\mathring{I_1}\= I + J^{-2}$ and $\mathring{I_2}\=J^2+IJ^{-2}$ are expressed in terms of $I\=\tr(\bm{FF}^T)$ and $J\=\hbox{det}(\bm{F})$~\cite{knowles1983large}, where $\bm{F}$ is the 2D (in-plane)  deformation gradient~\cite{holzapfel}. The dimensionless parameter $\epsilon$ quantifies the deviation of `strongly' nonlinear elasticity from the nonlinearity inherent in neo-Hookean elasticity (the first term in Eq.~\eqref{eq:Energy_den}).
\begin{figure}[h]
 	\includegraphics[width=1\linewidth]{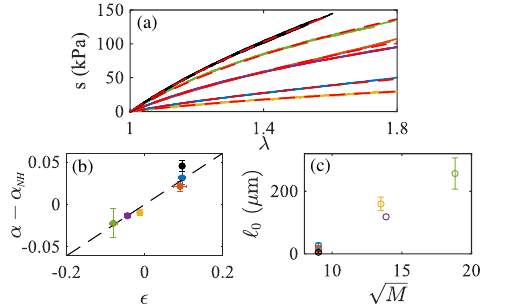}
 \caption{(a) Tensile nominal stress $s$ vs.~stretch $\lambda$ for the different gels (colors as in Fig.~\ref{fig:fig2}). The red dashed lines correspond to Eq.~\eqref{eq:Energy_den}, with $\epsilon\!=\!0.0965\pm0.0088$ (red), $0.0871\pm0.0167$ (blue), $0.0861\pm0.0016$ (black), $-0.0121\pm0.0068$ (yellow), $-0.0437\pm0.0099$ (purple), $-0.0819\pm0.0017$ (green). (b) $\alpha-\alpha_{_{\rm NH}}$ vs.~$\epsilon$, $\alpha_{_{\rm NH}}$ is for a purely neo-Hookean material (Eq. \ref{eq:h_lnl}). Dashed line is the best linear fit. (c) The shift $\ell_0$, defined in Eq.~\eqref{eq:h_lnl}, as a function of $\sqrt{M}$.}
 \label{fig:fig4}
\end{figure}

In Fig.~\ref{fig:fig4}a, we show that Eq.~\eqref{eq:Energy_den} perfectly describes the uniaxial tension stress-strain response of each gel at large $\lambda$, with deviations from neo-Hookian elasticity, $-0.0819<\epsilon <0.0965$, that  are only apparent at large $\lambda$. Figure~\ref{fig:fig4}b shows that the small variations in $\alpha$ are linearly dependent on $\epsilon$, when comparing $\alpha$ to $\alpha_{_{\rm NH}}$, the value of $\alpha$ for a purely neo-Hookean material, where $\epsilon\=0$. $h_{\rm max}$ is, therefore, related to {\it strongly} nonlinear elastic behavior, provided by both neo-Hookean nonlinearity and its strongly nonlinear correction, as quantified by $\epsilon$.

In contrast to $\alpha$, the shift in Eq.~\eqref{eq:h_lnl}, $\ell_0$, {\it is} a material-dependent quantity. What is its physical origin? Despite widely different values of  $\Gamma(v)$ and $\mu$ in the gels corresponding to the blue, orange and black symbols in Fig.~\ref{fig:fig3}c, their values of $\ell_0$ are nearly identical. These gels have the same value of $M$, suggesting that $\ell_0$ depends on $M$. Figure~\ref{fig:fig4}c indeed shows that $\ell_0$ is linearly related to $\sqrt{M}$. The value of $M$ corresponds to the average number of monomers between sequential crosslinks. The folded length between crosslinks, therefore, scales as $\sqrt{M}$~\cite{yang2019polyacrylamide,wang2023polyacrylamide}. This indicates that $\ell_{0}$ depends on a molecular length scale related to the polymer mesh size. As the Lake-Thomas model predicts that polymer chain scission occurs at scales $\xi\!\propto\!\sqrt{M}$~\cite{lake1967strength}, the linear relation between $\ell_0$ and $\sqrt{M}$ suggests that, in terms of fracture, $\ell_0$ is related to the dissipation length $\xi$.

\begin{figure}[h]
	\includegraphics[width=1\linewidth]{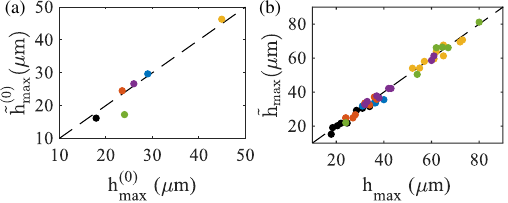}
	\caption{(a) Predictions for minimal $h_{\rm max}^{(0)}$, $\tilde{h}_{\rm max}(v\!=\!0)$, vs. $h_{\rm max}$ (measured at $v\!\simeq\!10^{-5}c_{_{\rm R}}$), see text for definitions and discussion. The dashed line of slope unity is a guide. (b) The predicted $\tilde{h}_{\rm max}$ vs.~the measured $h_{\rm max}$, see text for details. The dashed line highlights a perfect agreement.}
\label{fig:fig5}
\end{figure}

What determines $h_{\rm max}^{(0)}$, the minimal value of $h_{\rm max}$, in each gel? $h_{\rm max}^{(0)}$ defines the minimum length for which front interactions are stabilized. Equation~\eqref{eq:h_lnl} predicts that the minimum measured values, $h_{\rm max}^{(0)}$, correspond to $\tilde{h}_{\rm max}^{(0)}\!\equiv\!\alpha(\Gamma_0/\mu - \ell_0)$, where $\Gamma_0\=\Gamma(v\!\to\!0)$. This prediction is verified in Fig.~\ref{fig:fig5}a.

The above physical picture of step size selection implies that we can use Eq.~\eqref{eq:h_lnl}, together with energy balance, to predict $h_{\rm max}$ and $v_{\rm min}$ as follows. The total energy dissipation along the crack front is given by $\int_{0}^{w}\Gamma(v)\,dz\!\simeq\!\Gamma(v)(w + 1.4h)$~\cite{wang2022hidden}, which equals $G w$, where $w$ is the sample thickness and $G$ the measured energy flux into the crack. Substituting $\Gamma(v)$ above into Eq.~\eqref{eq:h_lnl}, and using energy balance, yield $h_{\rm max}$ and $v_{\rm min}$. Denoting the prediction for the former as $\tilde{h}_{\rm max}$, Fig.~\ref{fig:fig5}b shows that the measured $h_{\rm max}$  are in perfect agreement with this prediction.

To sum up, we have shown that interacting fronts in 3D select a well-defined and stable separation distance, the step size $h_{\rm max}$. This interaction is governed by two intrinsic lengthscales, a nonlinear elastic scale, $\ell_{\rm nl}\!\sim\!\Gamma(v)/\mu$, and a length $\ell_0$ that, like the dissipation scale $\xi$, depends on $\sqrt{M}$~\cite{lake1967strength}. Interestingly, recent experiments have shown that a scale proportional to $\sqrt{M}$ also governs a crack's transition to states propagating beyond the shear wave speed~\cite{wang2023tensile}. $\ell_{\rm nl}$ has been previously shown to play key roles in determining the shape (CTOD) of simple cracks~\cite{livne.08,livne2010,bouchbinder2010dynamics,bouchbinder2008weakly,bouchbinder20091,bouchbinder2014dynamics,boue2015failing} and in governing the transition to oscillatory motion of rapid simple cracks, where the oscillation wavelength's linear variation with both $\ell_{\rm nl}$ and $\xi$~\cite{lubomirsky2018universality,vasudevan2021oscillatory} is remarkably analogous to Eq.~\eqref{eq:h_lnl}.

Recent work has demonstrated that steps are initiated by material inhomogeneities of a characteristic scale~\cite{lubomirsky2024facet}. Here, we have shown that the selection of a unique and stable topology that characterizes steps is independent of how they are triggered and solely determined by the nonlinear interactions of fracture fronts in different planes. The stabilization of these entities is critically linked to nonlinear elasticity in soft brittle materials. The growth, decay and eventual stabilization of steps are determined by $\ell_{\rm nl}$, together with global energy balance. This discovery is an important step towards obtaining a fundamental understanding of crack front interactions in 3D systems, a largely open question. Additional work, e.g., regarding step drifting~\cite{vasudevan2020configurational} and step-step interactions~\cite{steinhardt2023geometric}, is required. To this end, a 3D fracture framework incorporating the intrinsic scales discussed here should be developed.

{\em Acknowledgements}. The authors thank the participants of a CECAM workshop on ``3D cracks and crack stability'' for enlightening discussions about crack segmentation. We also thank Songlin Shi and Mokhtar Adda-Bedia for helpful discussions, and Yuri Lubomirsky for useful comments on the manuscript. J.F. and M.W.~acknowledge the support of the Israel Science foundation (grant 416/23). E.B.~acknowledges support from the United States-Israel Binational Science Foundation (BSF, grant 2018603), the Ben May Center for Chemical Theory and Computation, and the Harold Perlman Family.

\clearpage

\onecolumngrid
\begin{center}
              \textbf{\Large Supplementary materials}
\end{center}

\setcounter{equation}{0}
\setcounter{figure}{0}
\setcounter{section}{0}
\setcounter{subsection}{0}
\setcounter{table}{0}
\setcounter{page}{1}
\makeatletter
\renewcommand{\theequation}{S\arabic{equation}}
\renewcommand{\thefigure}{S\arabic{figure}}
\renewcommand{\thesection}{S-\Roman{section}}
\renewcommand{\thesubsection}{S-\Roman{subsection}}
\renewcommand*{\thepage}{S\arabic{page}}
\renewcommand{\thetable}{S-\arabic{table}}

\section{Gel preparation}
Our experiments were performed on polyacrylamide gels, composed of acrylamide monomer chains crosslinked with bis-acrylamide crosslinker. The elastic properties of the gels are controlled by the concentration of the monomer and crosslinker. We prepared six different gels with different concentrations (w/v) of monomer. The monomer-to-crosslinker molar ratio, denoted as $M$, was the same for three of these. The polymerization was initiated by adding ammonium persulfate (APS) with a fixed concentration of 0.2$\%$ (w/v) and catalyzed with tetramethylethylenediamine (TEMED) with a fixed concentration of 0.05$\%$ (w/v). Table~\ref{tab:gel_composition} provides details of the gels composition (monomer and crosslinker concentrations) and the measured shear modulus $\mu$ (here, unlike in the table in Fig.~2 in the main text, we add the measurement error bars). Colors correspond to the gel labels in the main text. For gels with a fixed $M$, $\mu$ increases with the total polymer concentration $\rho_{\rm g}$ (see gels marked in blue, black and red in Table~\ref{tab:gel_composition}, and its caption to infer the values of $\rho_{\rm g}$). Moreover, $\mu$ decreases with decreasing crosslinker concentration under a fixed monomer concentration (see gels marked in blue and orange in Table~\ref{tab:gel_composition}).
\begin{table}[ht!]
\centering
\begin{tabular}{|c|c|c|c|c|c|}
 \hline
    & Gel & Monomer $\%$ & Crosslinker $\%$ & M\,\,\, & $\mu$\,(kPa)\,\,\, \\
    \hline
     \hline
  $\color{RoyalBlue}{\blacksquare}$ & 1 & 13.44 & 0.36 & 81.25 & 29.4 $\pm$ 1.5 \\
  \hline
  $\color{Black}{\blacksquare}$  & 2 & 29.22 & 0.78 & 81.25 & 110.4 $\pm$ 1.9 \\
  \hline
  $\color{Dandelion}{\blacksquare}$  & 3 & 13.44 & 0.16 & 183.19 & 20.2 $\pm$ 0.5 \\
  \hline
  $\color{BrickRed}{\blacksquare}$  & 4 & 21.04 & 0.56 & 81.49 & 63.1 $\pm$ 1.6 \\
  \hline
  $\color{Plum}{\blacksquare}$  & 5 & 27.87 & 0.3  & 192.56 & 67.1 $\pm$ 0.3 \\
  \hline
  $\color{LimeGreen}{\blacksquare}$  & 6 & 40.34 & 0.25 & 353.94 & 99.7 $\pm$ 3.9 \\
 \hline
\end{tabular}
\caption{A summary of the gels composition (monomer and crosslinker concentrations) and shear modulus $\mu$. Colors correspond to the gel labels in the main text. The total polymer concentration $\rho_{\rm g}$, defined in the main text, is the sum of the monomer and crosslinker concentrations appearing in the table.}
\label{tab:gel_composition}
\end{table}

\section{Velocity-dependent fracture energy and $h_{\rm max}$}

We measured the fracture energy of each gel using the crack tip opening displacement (CTOD) of the simple crack, as detailed in~\cite{boue2015failing}. As shown the main text, the fracture energy obeys a power-law dependence on the crack speed, $v$, according to $\Gamma(v)\!=\!\Gamma_0[1 + a(v/c_{_{\rm R}})^b]$. The values of $\Gamma_0$ and dimensionless parameters $a$ and $b$ obtained through the best fit of measured data are presented in Table.~\ref{tab:fit_parameters}. It is interesting to note that gels with a fixed monomer concentration and decreasing crosslinker concentration feature an increasing fracture energy (see gels marked in blue and orange in Table~\ref{tab:gel_composition} for the monomer and crosslinker concentrations, and Table.~\ref{tab:fit_parameters} for the $\Gamma_0$ values). This is in line with recent experiments showing that polymer networks with a longer chain length, corresponding in our context here to a fixed monomer concentration and decreasing crosslinker concentration, give rise to gels with higher fracture energy~\cite{hassan2022polyacrylamide}. In addition, the relation between the asymptotic step height $h_{\rm max}$ and the asymptotic crack speed $v_{\rm min}/c_{_{\rm R}}$ also shows a well-defined power-law. The best fit of $h_{\rm max}\!-\!v_{\rm min}/c_{_{\rm R}}$ using a power-law is highlighted by the black dashed curve in Fig.~3b in the main text.
In Table.~\ref{tab:fit_parameters}, we present the exponent of such power-law dependence (denoted as $c$ therein). The value of $c$ is very close to $b$, in agreement with the linear dependence of $h_{\rm max}$ on $\Gamma(v)/\mu$.
\begin{table}[ht!]
\centering
\begin{tabular}{|c|c|c|c|c|c|}
 \hline
    & Gel & $\Gamma_0$ &\,\,\, $a$\,\,\, &\,\,\, $b$\,\,\, &\,\,\, $c$\,\,\, \\
    \hline
    \hline
  $\color{RoyalBlue}{\blacksquare}$ &\,\, 1 \,\, &\,\, 5.53 \,\,&\,\, 3.46\,\, & \,\,0.624\,\, & \,\,0.625\,\, \\
  \hline
  $\color{Black}{\blacksquare}$ & 2 & 8.64 & 8.99 & 0.586 & 0.601  \\
  \hline
  $\color{Dandelion}{\blacksquare}$ & 3 & 9.51 & 4.19 & 0.569 & 0.627  \\
  \hline
  $\color{BrickRed}{\blacksquare}$ & 4 & 9.70 & 5.94 & 0.645 & 0.671 \\
  \hline
  $\color{Plum}{\blacksquare}$ & 5 & 19.9 & 8.02 & 0.573 & 0.522  \\
  \hline
  $\color{LimeGreen}{\blacksquare}$ & 6 & 37.7  & 26.55 & 0.477 & 0.431 \\
 \hline
\end{tabular}
\caption{The static fracture energy $\Gamma_0$ and fitting parameters $a$, $b$ and $c$ (see text for details) for the different gels.}
\label{tab:fit_parameters}
\end{table}

\section{Fracture experiments on particle-embedded gels}

We tested the effect of material heterogeneities on the scaling relation between the asymptotic step height, $h_{\rm max}$, and $\Gamma(v)/\mu$. To this end, additional experiments were performed using gels with 13.44$\%$/0.36$\%$ (blue in Table.~\ref{tab:gel_composition}) and 13.44$\%$/0.16$\%$ (orange in Table.~\ref{tab:gel_composition}) monomer/crosslinker concentration (w/v). We mixed a small amount of polyamide particles with a diameter of $\sim\!50\,\mu$m in the polymer solution. These particles, having a stiffness of a few GPa, can be considered as `infinitely rigid' inclusions in the gels~\cite{rozen2020fast}. Upon polymerization of the gels, the particles are randomly distributed in the gel sample with an average number density of 4-12 particles per mm$^2$.

\begin{figure}[ht!]
\centering
	\includegraphics[width=0.8\linewidth]{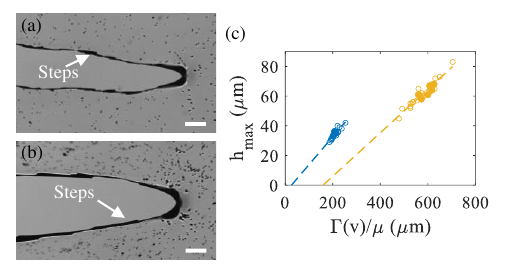}
\caption{(a) and (b) present snapshots of typical crack propagation in particle-embedded gels composed of monomer/crosslinker concentrations 13.44$\%$/0.36$\%$ and 13.44$\%$/0.16$\%$, respectively. Examples of step left on the crack surface are marked by the white arrows. Scale bar: 1 mm. (c) $h_{\rm max}$ as a function of $\Gamma(v)/\mu$ for the gels marked in blue and orange as in the main text. Solid symbols: step height measured in gels with embedded particles. Open symbols: step heights in the same gels with no embedded particles. The dashed lines correspond to the best linear fits.}
\label{fig:figS1}
\end{figure}

The fracture experiments are as described in the main text. Figures~\ref{fig:figS1}a-b show two snapshots of the crack propagation in the particle-embedded gels. Steps are intensively triggered in these gels, particularly when the crack meets particles. We consider cracks generating a single step. In Fig.~\ref{fig:figS1}c, we present the measured $h_{\rm max}$ versus $\Gamma(v)/\mu$ for steps both generated spontaneously (no embedded particles) and gels where the steps were triggered via particle interactions. The measurements in gels with particles embedded are highlighted by the solid symbols. The comparison clearly shows that while local heterogeneities trigger step formation~\cite{lubomirsky2024facet}, they have no effect on the subsequent evolution of the steps~\cite{steinhardt2022material}.

\twocolumngrid

%

\end{document}